\documentclass[sigconf]{acmart}
\usepackage[english]{babel}
\usepackage{blindtext}
\usepackage{lipsum} 
\usepackage{amsmath}
\usepackage{algorithm}
\usepackage[noend]{algpseudocode}

\usepackage{ftnxtra} 
\usepackage{fnpos}
\usepackage{balance} 

\usepackage{listings}
\usepackage{xspace}
\usepackage{listings}
\usepackage{cleveref}
\usepackage{multirow}
\usepackage{subfigure}
\usepackage{enumitem}
\newif\ifacm
\acmtrue

\newif\ifcomment
\commenttrue

\newif\ifcameraready
\camerareadyfalse

\newif\ifwatermark
\watermarkfalse

\ifwatermark
      \usepackage{draftwatermark}
      \SetWatermarkScale{.7}
      \SetWatermarkText{\shortstack{In Submission:\\Do Not Distribute}}
      \SetWatermarkColor[rgb]{1,0.88,0.88}
\fi

\ifcomment
    \newcounter{MVNumberOfComments}
    \stepcounter{MVNumberOfComments}
    \newcounter{YZNumberOfComments}
    \stepcounter{YZNumberOfComments}
    \newcounter{RCNumberOfComments}
    \stepcounter{RCNumberOfComments}

    \newcommand{\mvnote}[1]{\textcolor{magenta}{\small \bf [MV\#\arabic{MVNumberOfComments}\stepcounter{MVNumberOfComments}: #1]}}

    \newcommand{\yznote}[1]{\textcolor{red}{\small \bf [YZ\#\arabic{YZNumberOfComments}\stepcounter{YZNumberOfComments}: #1]}}
    
    \newcommand{\rcnote}[1]{\textcolor{orange}{\small \bf [RC\#\arabic{RCNumberOfComments}\stepcounter{RCNumberOfComments}: #1]}}

    \newcommand{\NOTE}[1]
    {
      {\footnotesize\it
        \begin{center}
          \begin{tabular}{|c|}
           \hline
            \parbox{0.85\columnwidth}{
              \medskip
              #1
              \medskip} \\
            \hline
          \end{tabular}
        \end{center}
        }
    }
\else
    \newcommand\mvnote[1]{}
    \newcommand\yznote[1]{}
    \newcommand\rcnote[1]{}
    \newcommand\NOTE[1]{}
\fi

\newcommand{\eg}{{e.g.,}\xspace}
\newcommand{\ie}{{\it i.e.,}\xspace}

\newcounter{NumTakeaways}
\stepcounter{NumTakeaways}

\newcommand{\numusers}{{four}\xspace}
\newcommand{\numcountries}{{six}\xspace}
\newcommand{\numdays}{{120}\xspace}
\newcommand{\numkm}{{9,378}\xspace}
\newcommand{\numcities}{{20}\xspace}
\newcommand{\findmy}{{\texttt{FindMy}}\xspace}
\newcommand{\smart}{{\texttt{SmartThings}}\xspace}

\definecolor{lightgray}{rgb}{.9,.9,.9}
\definecolor{darkgray}{rgb}{.4,.4,.4}
\definecolor{purple}{rgb}{0.65, 0.12, 0.82}

\lstdefinelanguage{JavaScript}{
  keywords={break, case, catch, continue, debugger, default, delete, do, else, false, finally, for, function, if, in, instanceof, new, null, return, switch, this, throw, true, try, typeof, var, void, while, with},
  morecomment=[l]{//},
  morecomment=[s]{/*}{*/},
  morestring=[b]',
  morestring=[b]",
  ndkeywords={class, export, boolean, throw, implements, import, this},
  keywordstyle=\color{blue}\bfseries,
  ndkeywordstyle=\color{darkgray}\bfseries,
  identifierstyle=\color{black},
  commentstyle=\color{purple}\ttfamily,
  stringstyle=\color{red}\ttfamily,
  sensitive=true
}

\begin{document}
\title{I Tag, You Tag, Everybody Tags!}

\author{Hazem Ibrahim}
\affiliation{%
  \institution{New York University Abu Dhabi}
  \city{Abu Dhabi}
  \country{UAE}
}
\email{hazem.ibrahim@nyu.edu}

\author{Rohail Asim}
\affiliation{%
  \institution{New York University Abu Dhabi}
  \city{Abu Dhabi}
  \country{UAE}
}
\email{rohail.asim@nyu.edu}

\author{Matteo Varvello}
\affiliation{%
  \institution{Nokia, Bell Labs}
  \city{New Jersey}
  \country{USA}
}
\email{matteo.varvello@nokia.com}

\author{Yasir Zaki}
\affiliation{%
  \institution{New York University Abu Dhabi}
  \city{Abu Dhabi}
  \country{UAE}
}
\email{yasir.zaki@nyu.edu}

\begin{abstract}
Location tags are designed to track personal belongings. Nevertheless, there has been anecdotal evidence that location tags are also misused to stalk people. Tracking is achieved \textit{locally}, \eg via Bluetooth with a paired phone, and \textit{remotely}, by piggybacking on location-reporting devices which come into proximity of a tag. This paper studies the performance of the two most popular location tags (Apple's AirTag and Samsung's SmartTag) through \textit{controlled} experiments -- with a known large distribution of location-reporting devices -- as well as \textit{in-the-wild} experiments -- with no control on the number and kind of reporting devices encountered, thus emulating real-life use-cases. We find that both tags achieve similar performance, \eg they are located 55\% of the times in about 10 minutes within a 100~m radius. It follows that real time stalking to a precise location via location tags is impractical, even when both tags are concurrently deployed which achieves comparable accuracy in half the time. Nevertheless, half of a victim's exact movements can be backtracked accurately (10m error) with just a one-hour delay, which is still perilous information in the possession of a stalker.

\end{abstract}





\maketitle

\section{Introduction}
\label{sec:intro}
Location tags such as AirTag (Apple) and SmartTag (Samsung) enable the monitoring of the location of any object they are attached to. This is achieved \textit{locally} by using Bluetooth Low Energy (BLE) -- or using Ultra Wideband if supported -- between a tag and the device it is paired with. When the location tag is out of reach, location updates are provided \textit{remotely} by piggybacking on any compatible iOS device, such as iPhones and iPads (for AirTag), or Samsung Galaxy devices (for SmartTag) which come into proximity of such tag. For a device to be \textit{location-reporting}, \ie eligible to relay a tag's location, it must support location finding, which Apple enables by default, but must be opted in on Samsung devices. 

Although the intended use case of tags is locating objects, there is anecdotal evidence of their misuse for tracking people~\cite{airtag_stalking, levitt_2022} or \textit{stalking}. To the best of our knowledge, no scientific study has yet quantified the accuracy of location tags in the wild, which directly correlates with their ability (or not) to act as stalking devices. Their efficacy in both locating and/or stalking depends on a few factors: 1) the technology adopted, and 2) the probability of encountering a location-reporting device, \eg a Samsung or Apple with enabled Bluetooth, GPS location, and data connectivity. While the reach of the technology can be studied in a lab, the opportunistic encountering of a reporting device requires experiments in the wild to account for realistic conditions. 

The goal of this paper us to study the performance of location tags. We tackle this problem with both \textit{controlled} and \textit{in-the-wild} experiments.  To enable such experiments, we develop crawlers for each tag's companion app (\findmy  and \smart) which collect fine-grained tags location histories as reported by location-reporting devices. We use controlled experiments to shed some light on the behavior of location tags, \eg how frequently their location is reported. First, we deploy an AirTag and a SmartTag in a secluded area along with Samsung and Apple devices at increasing distance. Next, we deploy both tags in our campus cafeteria whose WiFi provides us an estimate of the number of Apple and Samsung devices present at any point in time.

We use experiments in the wild to comment on the effectiveness of their opportunistic location reporting in various scenarios, \eg user mobility,  population densities, times of day, and days of the week. We rely on \numusers volunteers to carry an AirTag (Apple) and SmartTag (Samsung) while traveling to \numcountries different countries. The tags are mounted on the cover of an Android phone -- not paired with the tags, and not an Apple or Samsung --  which is equipped with a custom app logging information like GPS location, connectivity, etc. The data collected spans \numdays days, and \numkm Kms traveled across \numcities cities.

Our analysis shows that AirTag and SmartTag achieve similar performance with respect to how quickly and precisely they can be located. Despite the lower probability to encounter Samsung location-reporting devices, SmartTag matches AirTag performance by adopting a more \textit{aggressive} strategy of higher power Bluetooth beacons -- which are received further away but require 20\% more battery usage -- and more frequent location reports. Overall, 10 minutes are needed to locate a tag within 100~m from its true location, even when emulating a scenario where Apple devices can report SmartTags' location and vice-versa. This result implies that a victim's precise location can rarely be tracked in real time, even when both tags are concurrently deployed. Still, half of a victim's movements can be retraced with a 10 meters accuracy after just one hour, which is still dangerous information in the hand of a stalker. Furthermore, such stalking capability  is achieved with minimal cost (\$30 per tag) and for a very long time, given that both devices guarantee a battery life of about one year.


\section{Background and Related Work}
\label{sec:related}

Location tags like AirTag (Apple) and SmartTag (Samsung) use the Bluetooth Low Energy (BLE)~\cite{ble} protocol to transmit a unique identifier with a range of up to 100 meters. SmartTag+ and the AirTag also support Ultra Wideband~\cite{uwb} which further extends the range while allowing more precise device localization.  Ultra Wideband is only supported by recent devices, such as iPhone models $\geq$ 11, and Samsung Galaxy from the S21 onwards.

In addition to local (Bluetooth) tracking, remote tracking is achieved by allowing location-reporting devices -- iOS devices for AirTags, Samsung devices for SmartTags -- to report the location of tags encountered in the wild. Whenever a location-reporting device comes in the proximity of a location tag, \ie it receives a Bluetooth beacon, it updates the tag's location in the cloud using its GPS coordinates as an approximation. Tag owners can check their location via the tag's companion application. This process is private, without leaking any information about either the tag's owner or the device which has reported its last location. 

Apple and Samsung have implemented measures to deter malicious tracking, yet these measures have been insufficient, as discussed in~\cite{heinrich2021can}. In addition, each vendor only alerts a user if an unpaired tag from the same vendor has been in their vicinity for an extended period of time. This means that an AirTag could be used to stalk Samsung users and vice-versa. To address this, Apple released ``Tracker Detect''~\cite{tracker_detect} an Android application which allows its users to manually scan for nearby AirTags. Heinrich et al.~\cite{heinrich2022airguard} improved this design by automatically alerting users if they encounter the same AirTag in three separate locations within a 24 hour period. Similarly, Briggs et al.~\cite{briggs2022ble} extend the design proposed by Heinrich et al. to generic tags, not just AirTags. These applications are only partially effective due to MAC address randomization~\cite{mac_random}, which makes tags eventually appear as new devices to a third-party app. Mayberry et al.~\cite{mayberry2021tracks} developed a custom location tag which mimics an AirTag, can be tracked in Apple's \findmy network, and can circumvent Apple's tracking of malicious AirTags. Last but not least, Shaqfat et al.~\cite{shafqat2023track} also demonstrate that the security measures implemented by Apple can be circumvented with a custom location tag, and suggest solutions to mitigate stalking risks from cloned AirTags and enhance the existing anti-stalking safeguards for AirTags.

To the best of our knowledge, no previous paper investigates the performance (\ie accuracy and responsiveness) of tags in real-world scenarios on a global scale. Instead, Givehchian et al.~\cite{givehchian2022evaluating} have investigated the privacy of devices using the BLE protocol, such as location tags, showing that physical-layer identification is viable although often unreliable. Hernández et al.~\cite{hernandez2023analysis} have studied the efficiency of finding AirTags and Tile tags on a university campus through real and simulated experiments, where they model the probability of locating a tag utilizing the \textit{flow rate} of individuals carrying a compatible smartphone in the tag's detection range. They showed that AirTags exhibit a range between 10-30~m (inline with our own findings). Furthermore, their results indicate that given a populated area, both tags relayed their location within one hour 98\% of the time. Our work differs from the work of ~\cite{hernandez2023analysis} in that we expand our analysis to a global scale, analyzing the efficiency of AirTags and SmartTags across multiple countries. Moreover, we enhance their real experiments with a more comprehensive controlled experiment to evaluate the relationship between smartphone density and location update rate. 

\begin{figure}[t]
    \centering
    \vspace{-0.15in}
    \includegraphics[width=0.75\linewidth]{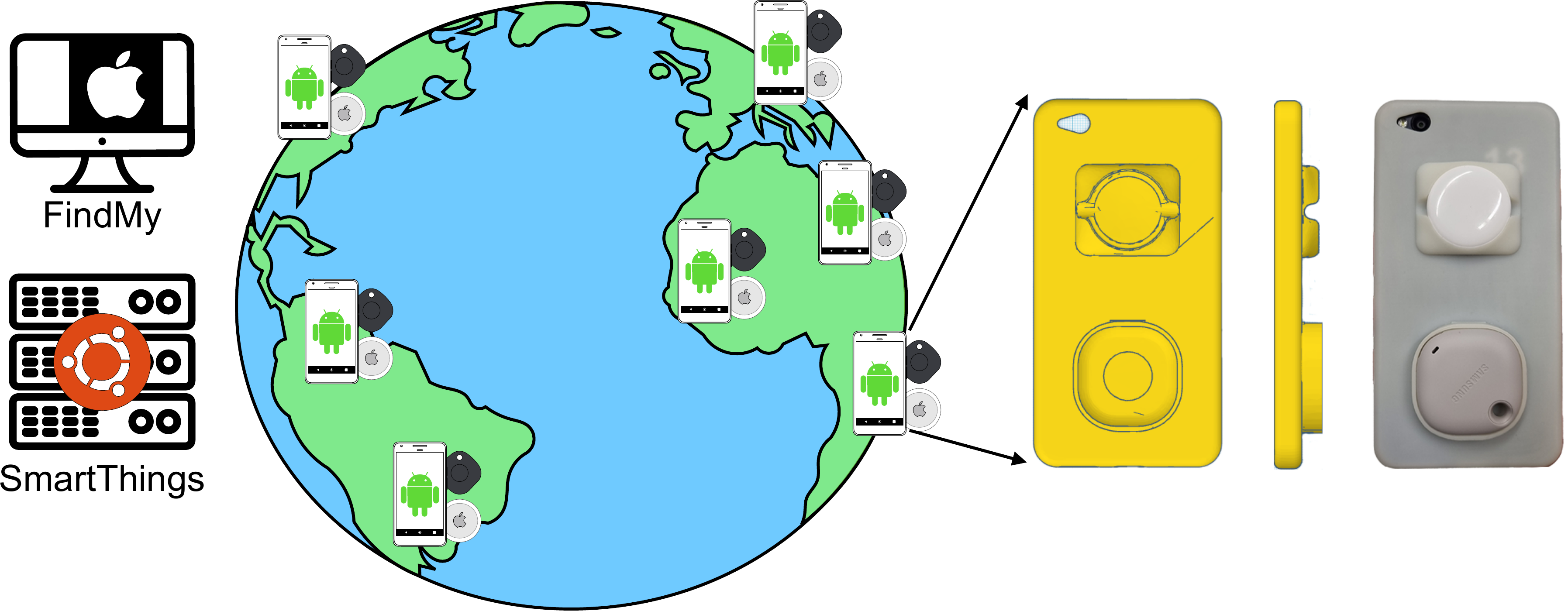}
    \caption{On the left, two data collection servers (MacOS and Ubuntu) run the \findmy and \smart crawlers. On the right, several views of our vantage point, a Redmi Go equipped with two tags.}
    \label{fig:arch}
    \vspace{-0.2in}
\end{figure}
\vspace{-0.05in}

\section{Methodology}
\label{sec:meth}
This section outlines the methodology we have devised to evaluate location tags. While we focus on AirTag (Apple) and SmartTag (Samsung), the methodology is generic and can be adopted to study other location tags like Tile~\cite{tile}.

\subsection{Location Tags Pairing}
\label{sec:sys:pairing}
\vspace{0.05in}
\noindent\textbf{Apple AirTag:} This tag must be paired and registered via Bluetooth with an iOS or iPadOS device above version 14.5, \ie no MacOS. Once the tag is linked to the Apple ID of the device it is paired with, it is then displayed in the \findmy app across all devices that have signed in with that Apple ID (including MacOS devices).

\vspace{0.05in}
\noindent\textbf{Samsung SmartTag:} This tag can only be paired and registered via Bluetooth with a Samsung Galaxy device running Android $\ge$8.0. The tag is linked to the Samsung account of the registered device, and it is displayed as a linked device in Samsung \smart app. 

\vspace{-0.1in}
\subsection{Tag Data Collection}
\label{sec:sys:data}
At the time when we developed our methodology, neither Samsung nor Apple offered public APIs to access tag's location data ($<$timestamp, GPS location$>$) as maintained by each tag's companion app: \findmy (Apple) and \smart (Samsung).  In addition, \findmy did not support location history, and \smart only provided some low resolution location history for up to 6 days. Therefore, we  developed ``crawlers'' for both apps which monitor location changes once a minute, and can thus build fine-grained location histories. Recently, Apple has released an API to allow authenticated users to query the \findmy service and retrieve location reports, as discussed in~\cite{bellon2023tagalong}.

\vspace{0.05in}
\noindent\textbf{\findmy Crawler:} The \findmy app is available for Apple devices, \eg Macbook, iPhone, and iPad. For ease of instrumentation, we write a crawler for MacOS. Note that MacOS v11 or above is needed, since \findmy on older MacOS does not support AirTags. In \findmy, users can find the last reported coordinates of any paired AirTag as follows. First, by clicking on the targeted tag from the list of devices in \findmy and selecting the option to open the location in Apple Maps. Once Apple Maps is launched, a pin is placed with the latest reported location of the tag. With a right-click, the user is given the option to ``copy coordinates''. 
We wrote a \findmy crawler in Python using \texttt{pyautogui}~\cite{pyautogui} to automate the above, and store the last reported coordinates of each available AirTag. Along with a tag's  coordinates, we also store a timestamp approximating when the coordinates were reported. This is computed using the crawling epoch time and the time at which a tag was last seen which is reported by \findmy as ``X minutes ago'', thus adding a potential error of up to one minute. Given this ``last seen'' time cannot be extracted from the \findmy app, we use OCR~\cite{OCR} to convert its value into text. 

\vspace{0.05in}
\noindent\textbf{\smart Crawler}: The \smart app is only available for Android. In the app, users select a tag from the list of tags associated with their account, and then click ``view location'' which opens Google Maps with a pin showing the tag's location. At this point, the tag's coordinates are available in the search bar and can be copied. We automate \smart via the Android Debugging Bridge (ADB~\cite{adb}), a rich Android protocol which allows to automate app operations like launching, scrolling, and GUI interaction. We connect an Android device, previously paired with one or more SmartTags, to a Linux machine via USB. ADB is then used to launch \smart and iterate over the tags. Once a tag's coordinates are available in Google Maps, they are copied and logged to a file. The same OCR-based procedure described for \findmy is used to approximate the time at which the tag location was updated last.  
\vspace{-0.1in}

\subsection{Vantage Point}
\label{sec:sys:remote}
A vantage point consists of an Android device (Xiaomi Redmi Go with a 1.4~GHz Quad-core and a 1~GB RAM), an AirTag and a SmartTag; both tags are mounted on a custom cover for the mobile device which we designed and 3D printed as part of the AmiGo testbed ~\cite{amigo} (see Figure~\ref{fig:arch}). The tags are paired with testing Samsung and Apple accounts as described in Section~\ref{sec:sys:pairing}. Note that the Android device used is not capable of reporting the location of neither tags, thus not impacting the accuracy of the experiments.

The Android device is equipped with an app we developed which collects GPS data, if available. The app buffers pairs of $<$timestamp, GPS location$>$ with a 5-second frequency for up to five minutes; only GPS variations are recorded, thus avoiding redundant data. After five minutes, the buffered data is POSTed to a server in our lab, if a data connection is available. Otherwise, the data is kept in the buffer until a connection becomes eventually available. The $<$timestamp, GPS location$>$ pairs are used as the ground truth of where the tags were located at a given point in time. This allows us to evaluate the accuracy of a tag's location as shown by its companion app, \ie as reported by location-reporting devices opportunistically encountered by location tags.

\begin{table}[t]
\small
\begin{tabular}{|l|c|c|c|c|c|} 
\hline
\textbf{Ctry} & \textbf{\# of} & \textbf{\# Report} & \textbf{\# Report} & \textbf{Walk/Jog/} & \textbf{Days} \\
      & \textbf{cities}    &   \textbf{Samsung} & \textbf{Apple} & \textbf{Transit(km)}  &  \\
\hline
US  & 2  & 145   & 4,821  & 14/22/871       & 30  \\ 
\hline
IT   & 10 & 1,361 & 4,520  & 157/68/3,170    & 28  \\ 
\hline
AE  & 2  & 1,442 & 9,572  & 145/151/3,384   & 52  \\ 
\hline
PK   & 1  & 129   & 454    & 13/16/165       & 2   \\ 
\hline
CH   & 1  & 331   & 489    & 14/16/62        & 3   \\ 
\hline
DE   & 4  & 187   & 1,225   & 46/45/1,021    & 5   \\ 
\hline
\hline
Tot.  & 20 & 3,595 & 21,081 & 388/317/8,673  & 120 \\
\hline
\end{tabular}
\caption{Summary of data-set collected in the wild. \# Report refer to the number of times that tag locations were reported as ``Now'' in each companion app.}
\vspace{-0.15in}
\label{table:dataset}
\end{table}

\begin{figure}[t]
    \centering
    \includegraphics[width=0.88\linewidth]{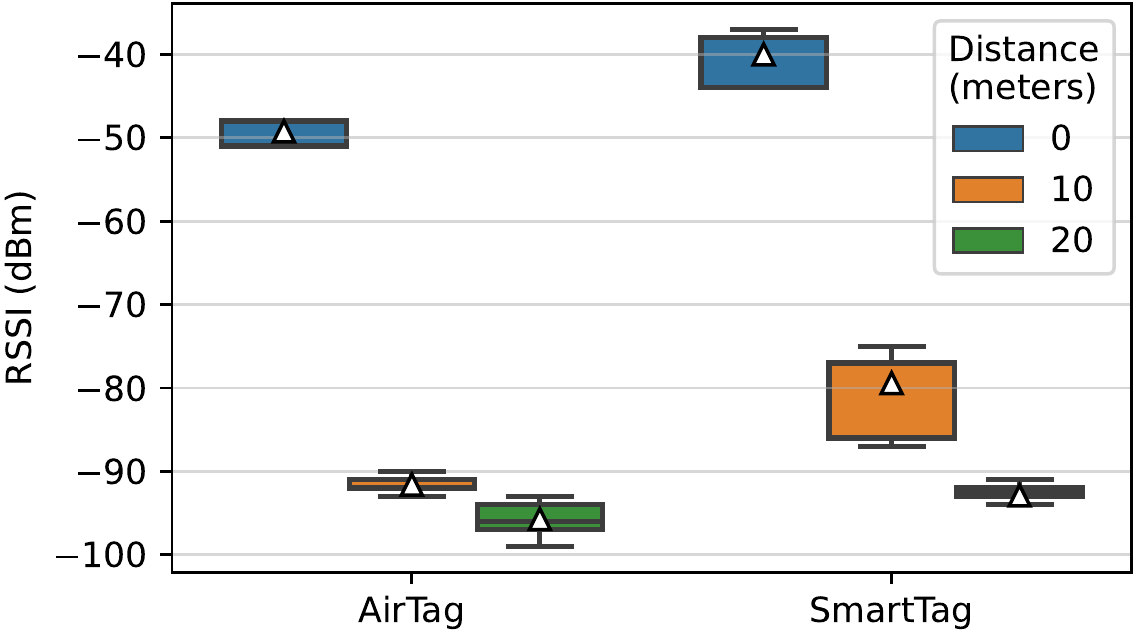}
    \vspace{-0.05in}
    \caption{Beacon RSSI for each tag at different distances.}
    \label{fig:beaconstrength}    
    \vspace{-0.1in}
\end{figure}
\begin{figure*}[t]
    \centering
    \includegraphics[width=0.9\linewidth]{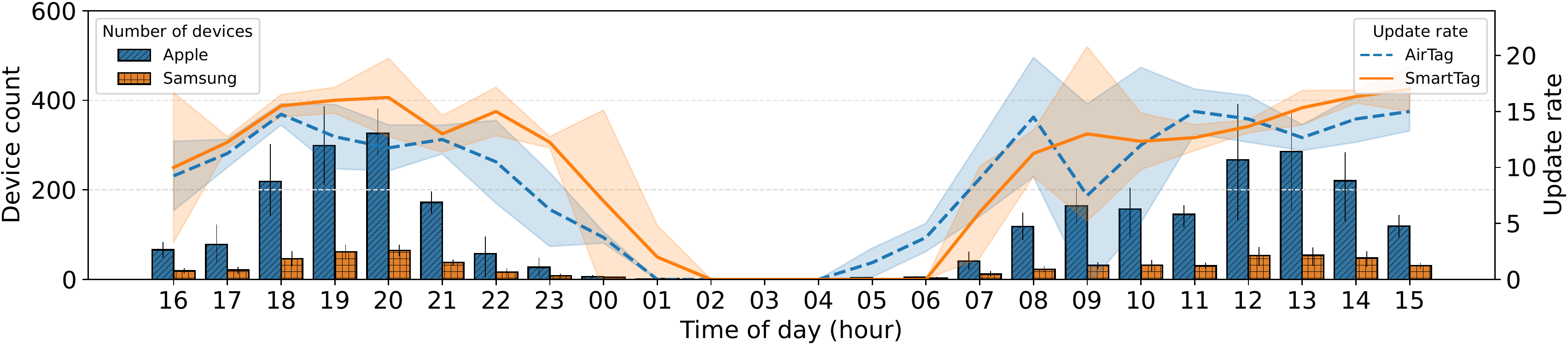}
    \vspace{-0.05in}
    \caption{Update rates of AirTag and SmartTag at different times of day in a busy university cafeteria.}
    \label{fig:controlled-updaterates}
\end{figure*}

\begin{figure*}[t]
    \centering
    \includegraphics[width=0.9\linewidth]{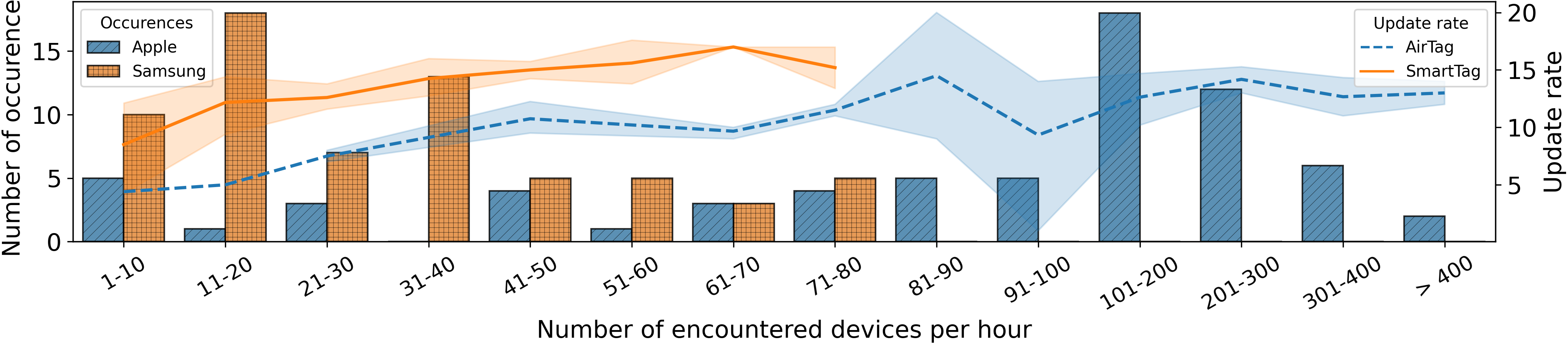}
    \vspace{-0.05in}    
    \caption{AirTag/SmartTag update rates as a function of the likelihood to have N reporting devices within one hour.}
    \label{fig:controlled-ratevscount}
\end{figure*}


\section{Data Collection and Analysis}
\label{sec:data}
This section describes two data-sets (controlled and in-the-wild) we have collected. It further details the crawling infrastructure used. Lastly, it introduces metrics and methodology we have devised to analyze location tag data-sets 

\vspace{0.05in}
\noindent\textbf{Controlled Experiments}  --  We deployed an AirTag and a SmartTag at a university cafeteria over five days. The cafeteria serves roughly 1,000 students, faculty, and staff, and operates everyday between 7:30am and 10pm, with peak hours during lunch (12 to 3pm) and dinner (6 to 9pm). Meanwhile, we ran our crawlers and collaborated with the university's IT infrastructure team to monitor the number of Apple and Samsung devices connected to the WiFi access point in the cafeteria. This is achieved by inspecting the destinations of the traffic generated by each device connected to WiFi. The rationale is that a clear distinction arises between Samsung and Apple devices since they rely on disjoint and proprietary data-centers to run their services. This was needed as modern mobile phones hide their vendor information from the MAC address~\cite{mac_random}. This information was aggregated into a count of the number of Apple and Samsung devices at different times, and thus completely anonymized. 

One limitation of this experiment is that we miss devices not connected to WiFi. While we cannot quantify this limitation, most phones rely on WiFi due to poor mobile coverage in the cafeteria. Another limitation is that we approximate the number of devices connected to WiFi to the number of reporting devices. This can be an overestimate, especially for Samsung devices whose users are required to opt-in to enable this behavior. 

We conduct an experiment in a secluded area -- 300~m away from any building -- where only our tags and phones are present. For each tag, we deploy four phones at distances of 0, 10, 20, and 50~m from the tag and measure both the frequency and strength of the Bluetooth beacons received. SmartTags beacons are easy to detect as they carry the sending tag name. Note that AirTags beacons can be uniquely identified since they share the first 4 bytes of their header (``1EFF004C12''). 

\vspace{0.05in}
\noindent\textbf{In-The-Wild Experiments} -- We deployed four vantage points (Android device reporting GPS location along with an AirTag and a SmartTag) via \numusers volunteers between March and August 2022. In total, the tags were carried along \numkm Kms across \numcountries countries and \numcities cities (see Table~\ref{table:dataset}). Participants were instructed to carry the vantage point as much as possible, and only interact with it to charge the phone, connect to WiFi, or insert a SIM card with a mobile data plan. 
To avoid biasing results in favor of either tag, participants ensured the location reporting option was disabled on any personal device they owned. Other family members were not required to do so. We filter data recorded within a 300~m radius of each participant's \textit{home}, as to not bias the data in the event of a neighbor or family member's phone repeatedly reporting a tag's location. Home locations are assumed as our participants homes, hotels, or any place they slept overnight. Overall, this filter accounted for 65\% of all data collected.

\begin{figure*}[tb]
    {\includegraphics[width=\linewidth]{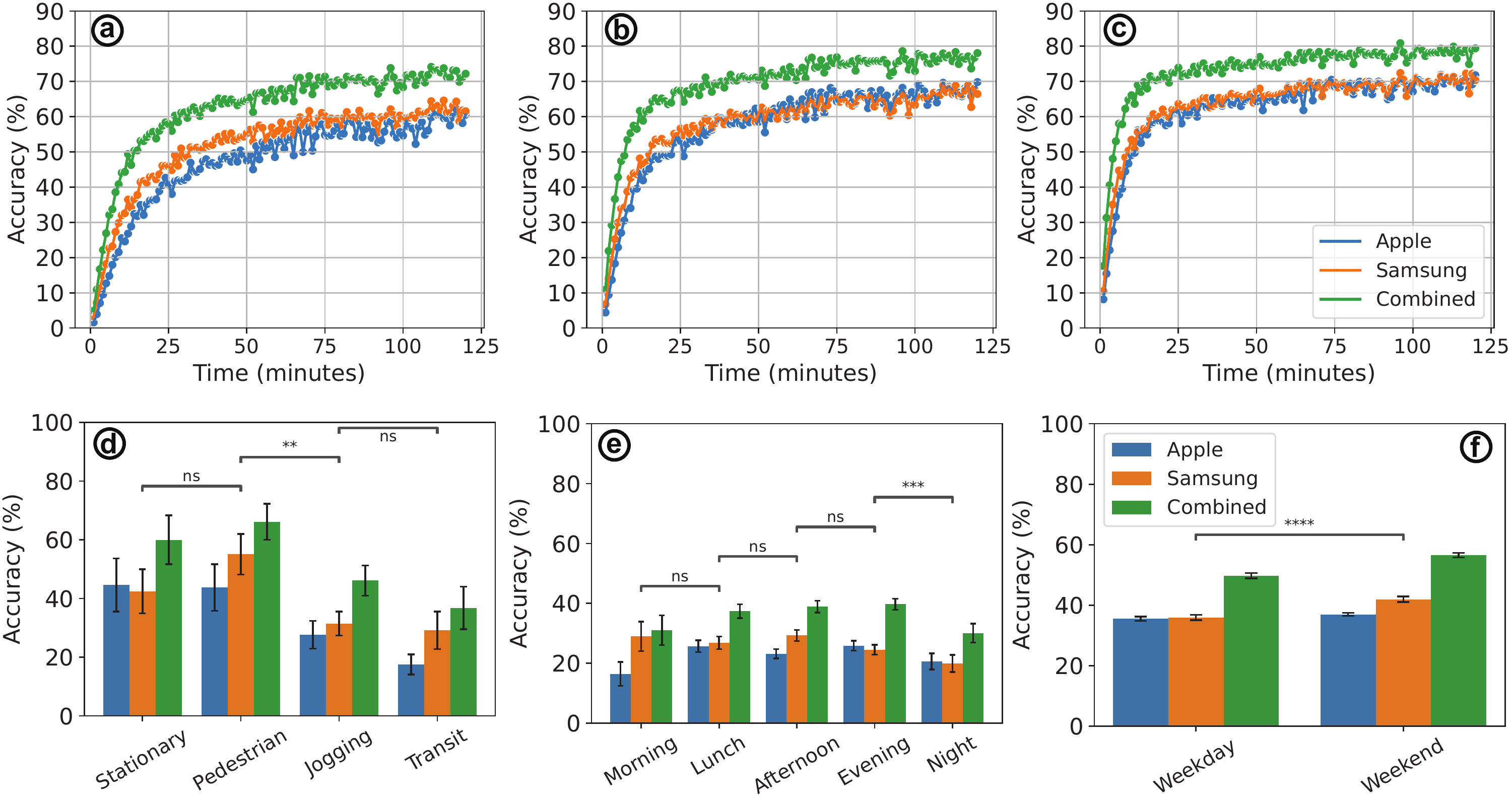}}
    \caption{Evaluation of AirTag, SmartTag and ``combined'' accuracy in the wild. (a) Time sweep (10 m radius). (b) Time sweep (25 m radius). (c) Time sweep (100 m radius). (d) Diff. speeds in a 10~min bucket. (e) Day times in a 10~min bucket. (f) Week days in a 10~min bucket.}
    \vspace{-0.15in}
\label{fig:percentagechancecontrolled}
\end{figure*}

We analyze the performance of AirTag and SmartTag both independently and \textit{combined}, which emulates a scenario where both devices report the location of each other's tags, functionally detaching the two tags from their proprietary ecosystems. This scenario is representative of a victim being stalked by both tags concurrently. We rely on two metrics: \textit{accuracy} and \textit{responsiveness}. 

\vspace{0.05in}
\noindent \textbf{Accuracy} -- At high level, assessing the accuracy of a tag consists of comparing its reported location, at a given time, with the location of its associated vantage point. Several factors might impact a tag's accuracy. First and foremost, the tag's location is approximated by the GPS location of the reporting device. Given Bluetooth has a 100 meter range, this can cause an error of up to 100~m. Another source of error is the movement of both the tag and the reporting device: as these devices move, the time needed to extract and report the GPS location can introduce some error. For example, when moving on a high speed train (300~Kmh) our sampling of the GPS locations every 5~s can introduce an error of up to 400~m. 

For a given tag, we group the locations reported  within the same X-minutes interval into the same ``bucket''. For each X-minutes ``bucket'', we calculate the distance between the location reported by the vantage point and the locations crawled from the tag's companion app. If the distance between the vantage point's location and a tag's location is below a (radius) threshold we count a ``hit'', otherwise we count a ``miss''. We compute a tag's accuracy as percentage of hits. 
To identify the radii of interest, we analyzed the combined accuracy of the location tags as we increase the radius of reporting across different time intervals (see Figure~\ref{fig:radiisweep} in Appendix~\ref{sec:app:radii}). In the case of short time intervals (1 and 10 minutes), the accuracy increases as the radius increases, eventually plateauing at roughly 100~m. For longer time intervals, there is no significant improvement in accuracy beyond 50~m. Accordingly, we will use the following radii in our analysis: 10, 25, and 100~m. 

\vspace{0.05in}
\noindent \textbf{Responsiveness}  --  Having accurate tag locations is important, but their locations also need to be reported in a timely manner. If a tag's location is updated frequently, then the owner will have less area to backtrack as (s)he realizes that the ``tagged'' object was lost. At the same time, a high update frequency is also an enabler of stalking or unsolicited tracking. We calculate tag responsiveness as the difference between the timestamp of the first hit -- \ie when the distance between the vantage point and a tag's location is below a radius --- and the first time that the vantage point reported such location.  

\vspace{0.05in}
\noindent \textbf{Ethics} -- 
We obtained IRB approval (HRPP-2021-185) and informed participants of our data collection practices through a consent form. While we collect GPS data, we do not gather any identifiable personal information.

\vspace{0.2in}
\section{Results}
\label{sec:res}

\subsection{Controlled}
\label{sec:res:control}

We start by analyzing the \textit{signal strength} of the Bluetooth beacons emitted by each tag. To do so, we measure the signal strength in a secluded area with phones. Figure~\ref{fig:beaconstrength} shows that, at a shorter distances (0 and 10 meters), SmartTag beacons are received with about 10dBm higher RSSI (Received Signal Strength Indicator) than AirTag beacons. However, at a distance of 20 meters, both tags' beacons were received at a similar RSSI.

Next, we investigate each tag's \textit{update rate}, computed as the number of location updates reported by location-reporting devices every hour. Accordingly, we focus on the controlled experiments performed in a cafeteria where the number of Samsung/Apple devices encountered by each tag naturally varies over time. Figure~\ref{fig:controlled-updaterates} shows the update rate as a function of the surrounding location-reporting devices. The figure shows, for each hour of the day, the average (over 5 days) tag's update rate and device count, \ie the number of Apple and Samsung devices present in the cafeteria. The shaded areas and error bars in the figure report the standard deviation of each metric. The figure shows an overall similar update rate between tags, peaking at roughly 15 updates per hour during lunch and dinner, and dipping to zero over night. However, the figure also shows that there were far more Apple than Samsung devices, up to 6 times more devices during peak hours, \eg 320 Apple devices versus only 50 Samsung devices at 8pm.  

To further understand the previous result, Figure~\ref{fig:controlled-ratevscount} shows the update rate as a function of the likelihood to have N location-reporting devices within one hour, \eg up to 10 and between 10 and 20. As expected from Figure~\ref{fig:controlled-updaterates}, it is more likely to find few Samsung devices, \eg less than 20, whereas it is more likely to find lots of Apple devices, \eg between 100 and 300. The key result of this analysis is that, while both AirTags and SmartTags converge to a similar maximum update rate (15-20 updates per hour), they do so in a very different way. Samsung implements an \textit{aggressive} update strategy, which quickly converges to the maximum update rate. In contrast, Apple implements a \textit{conservative} strategy, \eg half the update rate of Samsung when less than 20 devices are present. Samsung's update rate was not measured beyond 71-80 devices per hour due to the fact that there was never more than 80 Samsung phones in the cafeteria at any hour during the experiment.

\subsection{In-The-Wild}
\label{sec:res:wild}
\noindent
\textbf{Tags Accuracy and Responsiveness} -- We begin our analysis by investigating each tag's accuracy within a given radius as a function of its responsiveness. Figure~\ref{fig:percentagechancecontrolled} summarizes this analysis as we consider a radius of 10, 25, and 100 meters; note that ``combined'' refers to a unified Apple/Samsung ecosystem. Intuitively, Figure~\ref{fig:percentagechancecontrolled} (a,b,c) shows that relaxing the responsiveness, \ie allowing more time to locate a tag  within a radius, improves tag accuracy, \eg the combined tag's accuracy for larger radii (25  and 100 meters) grows from 10\% to 80\% as the responsiveness grows from one to 120 minutes. Combining tags offers a 15\% improvement, on average, over the accuracy of each individual tag. 

The previous observations also apply to a small radius (10 meters, see Figure~\ref{fig:percentagechancecontrolled}a) although with a few important differences. First, one minute is too fast to locate a tag within such a small radius, \eg an accuracy of 2\% versus 8-10\% at larger radii. Second, as we relax the responsiveness, the tag's accuracy increases much slower than what is observed for larger radii, \eg 40-45\% versus 60-63\% assuming a responsiveness of 25 minutes. This happens because, as both tags and reporting users might move, it is more challenging to correctly report the right location with such small radius and high responsiveness. Finally, the maximum accuracy caps at 72\%, when considering both tags combined, or 8\% less than what observed for larger radii.  Given the slow responsiveness allowed, this reflects errors introduced by approximating a tag's location with the reporting device location, which is unlikely more than 50 meters as per Figure~\ref{fig:beaconstrength}. 

Finally, if we focus on each tag independently, Figure~\ref{fig:percentagechancecontrolled}a shows that SmartTag slightly outperforms AirTag at a radius of 10~m but perform similarly at radii of 25 and 100~m, which is inline with Samsung's stronger signal strength at smaller radii, but similar performance at 20~m as shown in Figure~\ref{fig:beaconstrength}.

\vspace{0.05in}
\noindent
\textbf{Mobility and Time of the Day} -- We continue our analysis by exploring the effect of different mobility and temporal characteristics on the accuracy of each tag. For this analysis, we assume a responsiveness of 10 minutes and radii of 10, 25, and 100 meters. We also compute the statistical significance between different mobility and temporal scenarios by running t-tests across the average accuracy computed for each scenario. In Figures~\ref{fig:percentagechancecontrolled}d-f, statistical significant tests are denoted using the following symbols: ns denotes a $p>0.05$, * denotes $0.01<p<0.05$, ** denotes $0.001<p<0.01$, *** denotes $0.0001<p<0.001$, and **** denotes $p<0.0001$. 
 
Figure~\ref{fig:percentagechancecontrolled}d shows average tag's accuracy -- 95\% confidence intervals reported as error-bars across the different radii considered -- as we vary how fast a tag is moving (estimated as the average speed over 10 minutes as per our ground truth). We find that while walking at a pedestrian speed ($< 6.0$  km/h), the accuracy is maximized for both tags and even when combined. The rationale behind this finding is that walking represents a good equilibrium between number of devices the tag may be exposed to, \eg higher than when being stationary, and the length of the time window for the Bluetooth signal to be picked up by a location-reporting device. As the speed increases, \eg when jogging (speed comprised between $6.0$ and $12.0$  km/h) or in transit ($\geq 12.0$  km/h), the accuracy deteriorates due to the little time allowed for Bluetooth communication.

Figure~\ref{fig:percentagechancecontrolled}e shows a tag's mean accuracy during different times of the day. The figure shows no significant differences between morning (6 to 10 A.M.) lunch (10 A.M. and 2 P.M.), afternoon (2 to 6 P.M.) and evening hours (6 to 10 P.M.), but a statistically significant decrease at night (10 P.M. to 2 A.M.). Next, Figure~\ref{fig:percentagechancecontrolled}f shows significant tag's accuracy increase on weekends compared to weekdays, likely due to greater outdoor activity by the general public. 

\subsection{Discussion} 


This study highlights the effectiveness of AirTags and SmartTags in tracking the location of items or individuals across various situations. Specifically, the findings indicate that while these tags may not provide an immediate precise location, they can still offer an approximate location with just a one hour delay. This performance, although not flawless, carries significant implications for privacy and security, including the potential for misuse such as government surveillance of dissidents or domestic abuse-related stalking.

It is also important to acknowledge that this study is not without limitations.  While we are able to provide an estimate of the performance of location tags in real-world scenarios, this performance is ultimately dictated by the distribution of location-reporting devices in the vicinity of the tags, which we cannot quantify. Additionally, although our experiments aimed to encompass a diverse range of real-world scenarios, the study's scale was limited, involving only four participants carrying the tags and vantage points. Future research could expand upon these findings by involving a larger participant pool and exploring different geographical contexts.
\vspace{-0.1in}
\section{Conclusion}
Location tags such as AirTag and SmartTag are useful tools for locating objects, but there is anecdotal evidence of their misuse for tracking and stalking people. This paper has studied the performance of location tags through experiments in the wild and in controlled settings; all the data collected can be found at: \url{https://github.com/comnetsAD/Tags}. These experiments showed that AirTag and SmartTag achieve similar performance with respect to how quickly and precisely they can be located, in various scenarios. With respect to ``stalking'' people, both trackers (or a combination of the two) can effectively locate half of a victim’s  movements with 10 meters accuracy with just a one-hour delay. Conversely, real time stalking is less practical given that, most of the time, at least 10 minutes are required to achieve a 100 meters accuracy. It is important to note that these results might change over space, \eg in presence of different distributions of location reporting devices, and time, \eg due to future modification of the reporting protocols.


\bibliographystyle{ACM-Reference-Format}
\balance
\bibliography{references}


\appendix

\begin{figure}[tb]
    \centering
    \includegraphics[width=1\linewidth]{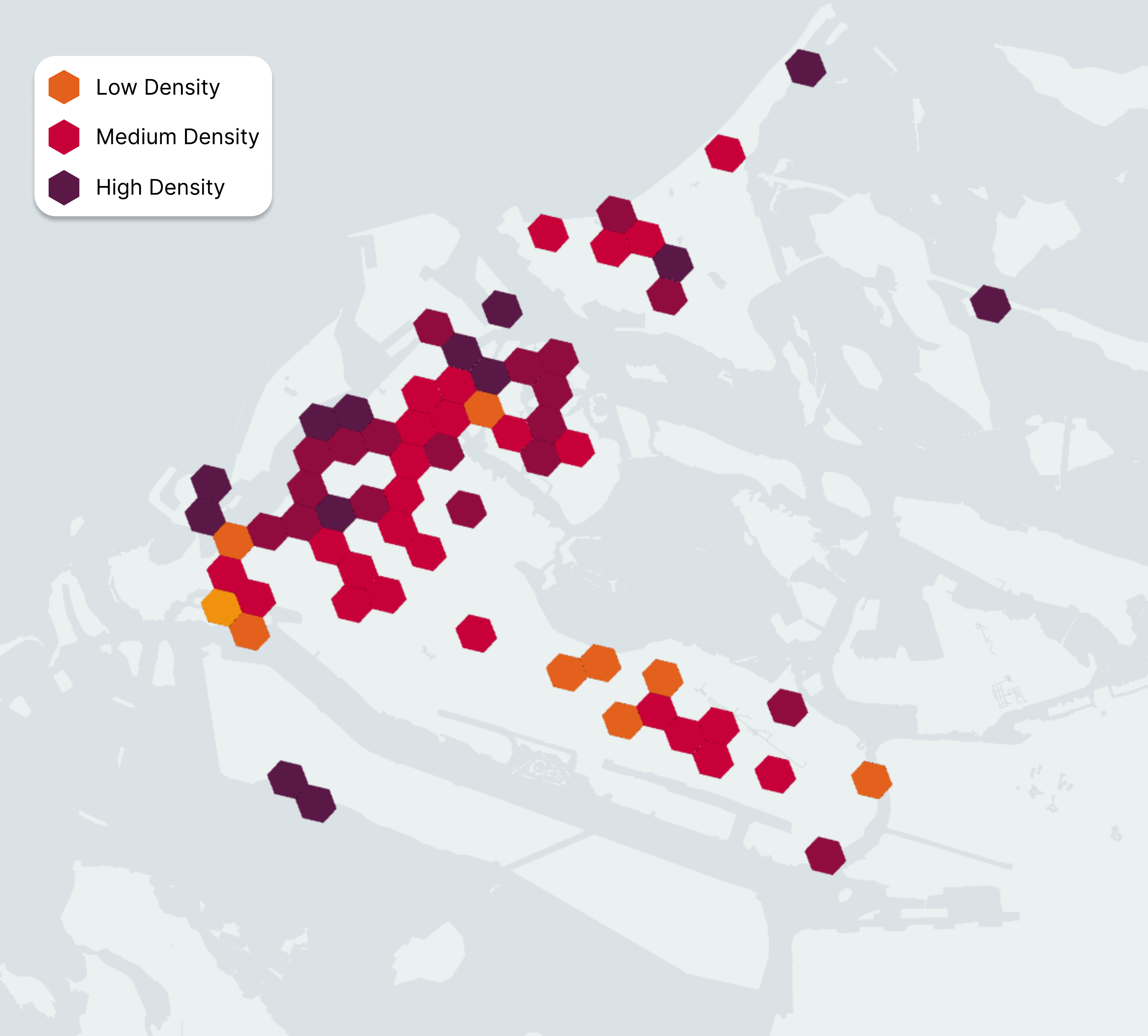}
    \caption{Visualization of hexagons visited by one of our study participants over 52 days in Abu Dhabi. Hexagons are derived assuming Uber's H3 index and a resolution of 8. The color of each hexagons relates to an estimate of population density provided by the Kontur data-set.}
    \label{fig:hexagons}
\end{figure}

\section{Hexagonal Hierarchical Spatial Index}
\label{sec:app:h3}
Uber's Hexagonal Hierarchical Spatial Index~\cite{uber} models the globe as an icosahedron, and creates 12 pentagons centered on each of its vertices joined by 110 hexagons. Each cell is then recursively filled with a number of hexagons which depends on the desired resolution. The total number of cells at a given resolution r is given by $c = 2 + 120*7^r$.  For instance, at a resolution of zero, the earth is covered by 110 hexagons, whereas at a resolution of eight, the number of hexagons increases to 691,776,110~\cite{h3_resolution}.  Naturally, as the number of cells which cover the surface of the earth increases, each cells occupies less area overall. At a resolution of eight, an individual hexagon has an average area of 0.737 km\textsuperscript{2}. In our analysis (see Section~\ref{sec:res:wild}), we use a resolution of eight as it coincides with the resolution used in the Kontur Hexagon Population density data set~\cite{kontur},  which reports population densities within H3 hexagons inferred from satellite images of building density. 

Figure~\ref{fig:hexagons} shows an example of the hexagons visited by one of our study participants in Abu Dhabi (UAE).  We consider an hexagon \textit{visited} if our participant (and tag) spent at least 5 consecutive minutes within it, thus ignoring hexagons which a user has only visited briefly, \eg while driving on the highway. We color code each hexagon using the Kontur data-set for population density, from low density (orange) to high density (dark read). Thresholds for classifying a particular hexagon as low, medium, or high density are explained in Section~\ref{sec:res:wild}.

\begin{figure}[b]
    \centering
    \includegraphics[width=1\linewidth]{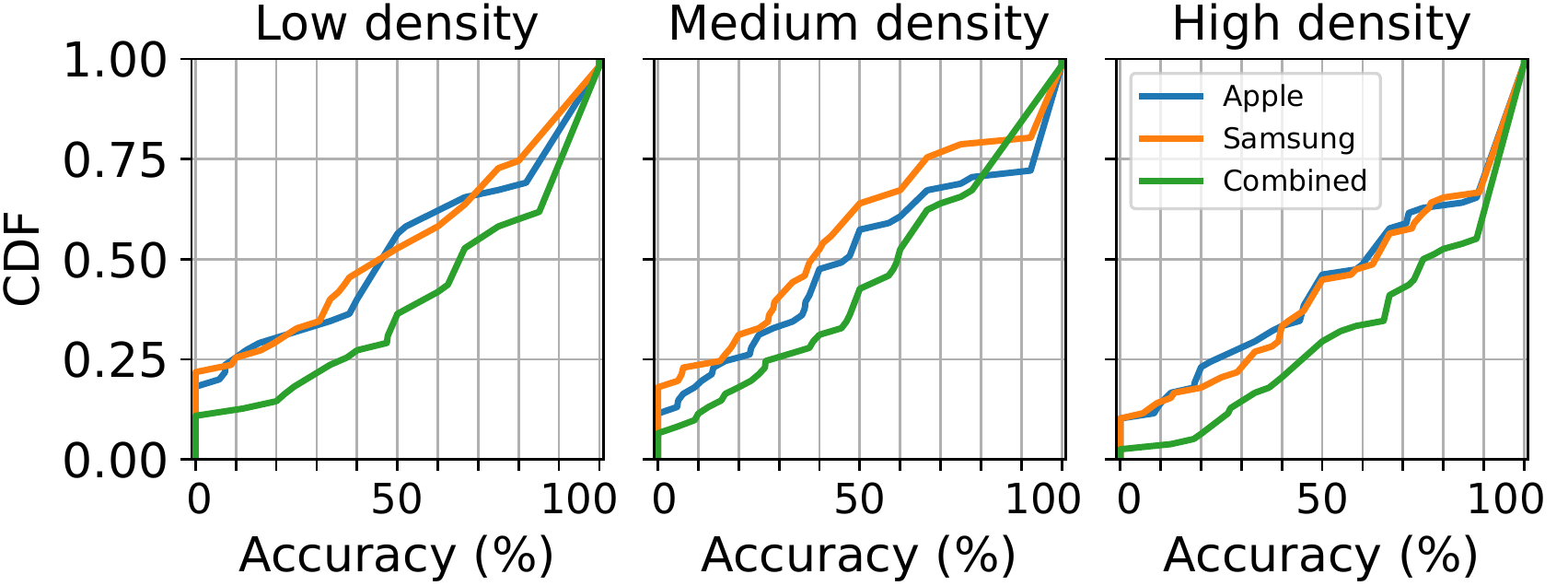}
    \vspace{-0.15in}
    \caption{CDF of tags accuracy for different population densities; one hour responsiveness and 100 meters radius.}
    \vspace{-0.2in}    
    \label{fig:densitycdf}
\end{figure}

\vspace{0.05in}
\noindent

\section{Impact of Population Density on Tag Accuracy}
\label{sec:app:popdensity}
Intuitively, the accuracy of a tag depends on the number and type of devices in their vicinity. While we cannot collect this information in the wild, we approximate it with the Kontur Hexagon Population density data set~\cite{kontur}, which reports population densities within H3 hexagons inferred from satellite images of building density. H3 is Uber's Hexagonal Hierarchical Spatial Index~\cite{uber} which groups GPS locations as hexagons.

We group GPS locations from our data-set as hexagons using a ``resolution'' of eight as in the Kontur data set; see Appendix~\ref{sec:app:h3} for more details. We threshold the different population density buckets as the 33rd, 66th, 100th percentiles of the population densities of all hexagons visited in our study. As such, we designate hexagons which hold a population $< 600$ (33rd percentile) as ``low density'', those with a $600 \le$ population $< 1,750$  (66th percentile) as ``medium denisty'' and those with population $\geq 1,750$ as ``high density''.   

Figure~\ref{fig:densitycdf} shows the Cumulative Distribution Function (CDF) of the accuracy as a function of population density (low, medium, and high). For this analysis, we consider a responsiveness of one hour and radius of 100 meters. The figure shows that the probability of a zero accuracy, \ie no correct location reported within 100 meters, drops from 20-25\% in low density areas down to 10-15\% in high density areas. A slight decrease in accuracy is observed between low and medium density areas for the median accuracy (roughly 45\% in low-density areas vs.\ 42\% in median-density areas), while it increases to 63\% for high density areas. With respect to the combined accuracy, high density areas see the least improvement: on average 15\% versus 20\% in low density areas. This happens because the benefit of sharing the same ecosystem reduces as an area is already highly populated with devices from each ecosystem.

\begin{figure}[tb]
    \centering
    \includegraphics[width=0.85\linewidth]{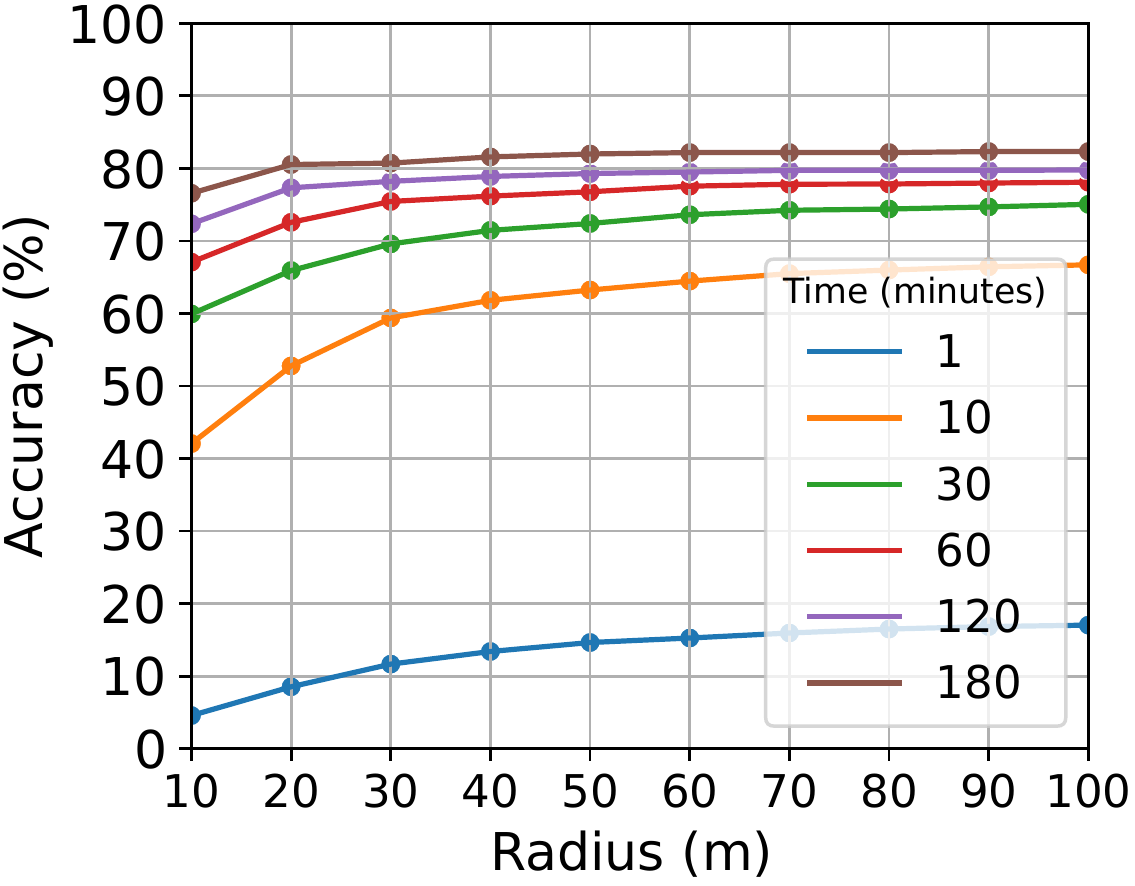}
    \caption{Combined accuracy of tags vs. radius across different time windows.}
    \label{fig:radiisweep}
\end{figure}

\section{Impact of Radii on Tag Accuracy}
\label{sec:app:radii}
To determine the \textit{accuracy} of a tag, we compare the vantage point's location (our ground truth) with a tag's location. If this distance is within a \textit{radius} we count a ``hit'', otherwise we count a ``miss''. We then compute a tag's accuracy as percentage of hits. To determine which radius values to use in our analysis, we have explored the impact of different radii on each tag's accuracy. Figure~\ref{fig:radiisweep} shows the combined tag's accuracy, \ie considering a unified Apple/Samsung ecosystem,  as a function of both radius and responsiveness, or how quickly a tag's location is correctly reported within a radius value.  We only show the combined tag's accuracy since its trend is representative of each tag's accuracy. When considering an aggressive responsiveness (1 minute), the accuracy greatly improves as the radius increases, starting with an accuracy of 5\% at 10 meters, and increasing to 17\% at 100 meters. When allowed a longer time to correctly locate a tag, the accuracy still increase as we increase the radius, but  plateaus at a smaller radius (between 50 and 70 meters) compared to a more aggressive responsiveness. Accordingly, for the purpose of our study we select  radii of 10, 25, and 100 meters.




\end{document}